\documentclass{aastex}

\def\apj{ApJ}                 
\def\apjl{ApJ}                
\def\apjs{ApJS}               
\def\aap{A\&A}                
\def\mnras{MNRAS}             
\def\nat{Nature}              

\usepackage{graphicx}

\shorttitle{BH spin-orbit misalignment in Galactic XRBs}
\shortauthors{T. Fragos et al.}

\begin{document}

\title{BLACK HOLE SPIN-ORBIT MISALIGNMENT IN GALACTIC X-RAY BINARIES} 

\author{T.\ Fragos$^{1}$, M.\ Tremmel$^{1}$, E.\ Rantsiou$^{2}$, K.\ Belczynski$^{3,4}$} 

\altaffiltext{1}{Department of Physics and Astronomy, Northwestern University, 2145 Sheridan Road, Evanston, IL 60208, USA}
\altaffiltext{2}{Department of Astrophysical Sciences, Princeton University, 4 Ivy Lane, Peyton Hall, Princeton, NJ 08544, USA}
\altaffiltext{3}{Oppenheimer Fellow, Los Alamos National Lab, P.O. Box 1663, MS 466, Los Alamos, NM 87545}
\altaffiltext{4}{Astronomical Observatory, University of Warsaw, Al. Ujazdowskie 4, 00-478 Warsaw, Poland}

\email{tassosfragos@northwestern.edu, michaeltremmel2007@u.northwestern.edu, emmarant@astro.princeton.edu, kbelczyn@nmsu.edu}

\begin{abstract}

In black hole X-ray binaries, a misalignment between the spin axis of the black hole and the orbital angular momentum can occur during the supernova explosion that forms the compact object. In this letter we present population synthesis models of Galactic black hole X-ray binaries, and study the probability density function of the misalignment angle, and its dependence on our model parameters. In our modeling, we also take into account the evolution of misalignment angle due to accretion of material onto the black hole during the X-ray binary phase. The major factor that sets the misalignment angle for X-ray binaries is the natal kick that the black hole may receive at its formation. However, large kicks tend to disrupt binaries, while small kicks allow the formation of XRBs and naturally select systems with small misalignment angles. Our calculations predict that the majority ($>67\%$) of Galactic field BH XRBs have rather small ($\lesssim 10^{o}$) misalignment angles, while some systems may reach misalignment angles as high as $\sim 90^{o}$ and even higher. This results is robust among all population synthesis models. The assumption of small small misalignment angles is extensively used to observationally estimate black hole spin magnitudes, and for the first time we are able to confirm this assumption using detailed population synthesis calculations.

\end{abstract}


\maketitle

\section{INTRODUCTION}

The formation of Roche lobe overflowing X-ray binaries (XRBs) with black hole (BH) accretors in galactic fields, involves the evolution of a primordial isolated binary of which at least one of the component stars is massive enough to form a BH. Right before the supernova (SN) explosion that will create the BH, the binary orbit is assumed to be circular, due to tidal interactions and possible preceding mass-transfer (MT) episodes, and the spins of the two stars are assumed to be aligned with the orbital angular momentum at the formation of the binary. The subsequent SN explosion will change the orbital characteristics of the binary. If the SN explosion is symmetric, the mass lost from the binary will cause the orbit to expand and induce some eccentricity, however the BH spin axis will remain aligned with the orbital angular momentum. In contrast, if an asymmetric kick is imparted on the BH during its formation, the orientation of the orbit will also change and this in turn will cause a misalignment between the spin axis of the BH and the orbital angular momentum.

Through studies of the position and spatial velocity of radio pulsars in the Galaxy, it has been established in the literature that neutron stars (NSs) can receive large asymmetric SN kicks ($\sim 265\rm Km\,s^{-1}$) \citep{GO1970, HP1997, Hobbs2005}. On the other hand, the possibility of an asymmetric kick imparted to BHs remains still an open question. Stellar BHs need to be in XRBs in order to be observable, and only for a handful of Galactic BH XRBs there are accurate measurements of their position and spatial velocity. However, indirect BH kick observations indicate that low mass BH receive high asymmetric kicks, while high mass BHs low or no kicks \citep{Mirabel2002, Mirabel2001, MR2003, Dhawan2007}. On the other hand, detailed theoretical modeling of the evolutionary history of two Galactic BH XRBs (GRO\,J1655-40 and XTE\,J1118+480) showed that for the case of GRO\,J1655-40  a moderate asymmetric kick is possible, while in the case of XTE\,J1118+480 an asymmetric kick larger than $\sim 100\rm Km\,s^{-1}$ is necessary in order to explain the current properties of the system \citep{Willems2005,Gualandris2005,Fragos2009}.

Unless the direction of this asymmetric kick during the formation of the BH is finely tuned to be on the orbital plane, it will induce a misalignment of the BH spin and the orbital angular momentum. In fact, two microquasar type Galactic BH XRBs (GRO\,J1655-40 and V4641 Sgr) have been observed with relativistic radio jets that are misaligned with respect to their orbital plane. This misalignment angle is estimated to be $\sim 15^{o}$ for GRO\,J1655-40 \citep{HR1995,OB1997}, and $\sim 55^{o}$ for V4641 Sgr \citep{Orosz2001}. Assuming that the radio jet has the same direction as the spin of the BH, this misalignment angle can be translated to a tilt of the BH spin axis with respect to the orbital plane. 


In this letter we present population synthesis (PS) models of Roche-lobe overflowing, Galactic field BH XRBs, and study their BH spin-orbit misalignment and its their dependence on model parameters. 

\section{X-RAY BINARY POPULATION SYNTHESIS MODELS}


We performed the simulations presented here with \textit{StarTrack} \citep{BKB2002, Belczynski2008}, a current binary PS code that has been tested and calibrated using detailed binary star calculations 
and incorporates all the important physical processes of binary evolution \citep[see ][ for a detailed description of all code elements]{Belczynski2008}. 


\subsection{Model Parameters and Initial Conditions for Milky-way Models}

We constructed a library of PS models appropriate for the Milky-way. We performed a parameter study varying some of the important parameters involved in the formation and evolution of XRBs, such as the IMF, the stellar wind strength for the late evolutionary stages of massive stars, the common envelope (CE) efficiency, the distribution of magnitude and direction of asymmetric kicks, and the initial spin magnitude of BHs. In all our models we adopted a solar metallicity for the stellar population, and a constant star formation rate over the age of the Galaxy, which is assumed to be $~10\rm \,Gyr$. 

For the full list of the model parameter combinations we considered, and the naming convention we used, see Table 1.  

\begin{deluxetable}{ccccc}
\tablecolumns{5}
\tabletypesize{\scriptsize}
\tablewidth{0pt}
\tablecaption{PS Models: For each of the models listed below we applied two types of asymmetric kicks, one with an isotropic distribution for the direction of the kicks (denoted by the letter \emph{``i''} after the model number), and one with polar kicks, where the kick is always directed perpendicular to the orbital plane (denoted by the letter \emph{``p''} after the model number). In addition, we considered different distribution for the initial BH spin magnitude: constant spin with value of 0.2, 0.5, or 0.9 across the whole BH population (denoted with an exponent \emph{``C0.2''}, \emph{``C0.5''}, or \emph{``C0.9''} correspondingly), or spin magnitudes in the range between 0 and 1 drawn from a Maxwellian distribution with a maximum at 0.2, 0.5, or 0.9 (denoted with an exponent \emph{``M0.2''}, \emph{``M0.5''}, or \emph{``M0.9''} correspondingly).
\label{models}} 
\tablehead{ \colhead{Model} & 
     \colhead{$\alpha_{\rm CE}$ \tablenotemark{a}} & 
     \colhead{IMF exponent} & 
     \colhead{$\eta_{\rm wind}$ \tablenotemark{b}} &
     \colhead{$\sigma\ V_{\rm kick}$ \tablenotemark{c}} 
     }
\startdata
1	& 0.5	& -2.35		& 1.0	& 265\\
2	& 0.3	& -2.35		& 1.0	& 265\\
3	& 0.7	& -2.35		& 1.0	& 265\\
4	& 0.5	& -2.35		& 1.0	& 50\\
5	& 0.5	& -2.35		& 1.0	& 150\\
6	& 0.5	& -2.35		& 1.0	& 400\\
7	& 0.5	& -2.7		& 1.0	& 265\\
8	& 0.5	& -2.7		& 1.0	& 50\\
9	& 0.5	& -2.7		& 1.0	& 150\\
10	& 0.5	& -2.7		& 1.0	& 400\\
11	& 0.5	& -2.35		& 0.25	& 265\\
12	& 0.5	& -2.35		& 0.25	& 50\\
13	& 0.5	& -2.35		& 0.25	& 150\\
14	& 0.5	& -2.35		& 0.25	& 400\\
15	& 0.5	& -2.7		& 0.25	& 265\\
16	& 0.5	& -2.7		& 0.25	& 50\\
17	& 0.5	& -2.7		& 0.25	& 150\\
18	& 0.5	& -2.7		& 0.25	& 400\\
19	& 0.5	& -2.35		& 1.0	& 265 \tablenotemark{d}\\

\enddata
\tablenotetext{a}{CE efficiency parameter}
\tablenotetext{b}{Stellar wind strength parameter for the late evolutionary stages of massive stars}
\tablenotetext{c}{$\sigma$ parameter of the Maxwellian distribution of asymmetric kick magnitudes}
\tablenotetext{d}{Asymmetric kicks magnitudes are drawn directly from the Maxwellian distribution and not scaled to the fall-back mass}

\end{deluxetable}

\subsection{Asymmetric Black Hole Kicks}

At the time of birth, BHs and NSs may receive an instantaneous asymmetric kick, due to asymmetries involved in the SN explosion mechanisms. The distribution of kick velocities for NSs has been estimated through studies of the position and spatial velocity of radio pulsars in the Galaxy. \citet{Hobbs2005} studied the proper motion of 233 Galactic pulsars and found that their inferred kick velocities are consistent with a single Maxwellian distribution with $\sigma=265\rm \, Km\, s^{-1}$. 

In the SN explosions of more massive stars that can form BHs, the effects of material fall-back (ejected initially in the SN explosion) during the star's final collapse are included. For each core-collapse event we calculate the fall-back factor ($f_{\rm fb}$), i.e., the fraction of the stellar envelope that falls back. The value of $f_{\rm fb}$ is calculate based on the mass of the carbon/oxygen core ($M_{\rm CO}$) of the pre-SN star, and is interpolated linearly between $M_{\rm CO}=5\, M_{\odot}$ ($f_{\rm fb}=0$) and $M_{\rm CO}=7.6\, M_{\odot}$ ($f_{\rm fb}=1$). The regimes of no fall-back, partial fall-back, and direct collapse are estimated from earlier studies by \citet{Fryer1999, FK2001}. 
In the formation of BHs, where partial fall-back or direct collapse takes place, kicks are lowered proportionally to the amount of fall-back associated with core-collapse event, according to the relation:
\begin{equation}
V_{\rm kick} = (1-f_{\rm fb})V,
\end{equation} 
where V is the kick magnitude drawn form the single Maxwellian. Furthermore, we consider the extreme case where the magnitude of the BH kick is drawn directly for the maxwellian distribution and not scale to fall-back mass (model 19). Finally, the standard assumption is that the direction of the kick is isotropically distributed, but we also consider the case of polar kicks, where the kick direction is aligned with the spin axis of the BH progenitor.

\subsection{Initial Black Hole Spin-Orbit Misalignment}

Here we assume that any asymmetry in the SN explosion starts at, or very close to, the center of the progenitor star. 
If this natal kick has a component perpendicular to the plane of the orbit, then the inclination of the system will change while the spin of the BH will remain virtually unaffected, causing the BH spin to no longer be aligned with the orbit. 

\emph{StarTrack} treats the orbital dynamics of the SN explosion by taking into account mass/angular momentum losses as well as asymmetries through natal kicks. The only simplifying assumption in these calculations is that the SN explosion is instantaneous \citep[for details see][]{Kalogera1996,Belczynski2008}. Knowing the exact geometry of the orbit right before and after the SN, and assuming that the direction of the BH spin axis remains unchanged relative to the spin axis of the BH progenitor, we are able to calculate the post-SN BH spin-orbit misalignment angle.

\section{BLACK HOLE SPIN EVOLUTION DUE TO ACCRETION}

During the XRB phase, the material that is being accreted onto the BH can be a significant fraction of the BH's initial mass, and the effects of the accretion on the BH spin can not be considered negligible. We therefore want to investigate whether this mass accretion can alter significantly the post-SN spin-orbit misalignment. In this section we describe how we model these effects.

Let us assume a BH of spin $a$ and mass $M_{\rm BH}$ surrounded by an accretion disk which has an inclination $i$ with respect to the BH's spin (the accetion disk is assumed to be parallel to the orbital plane), and that mass $m_{\rm acc}$ is gradually being accreted onto the BH. The mass accreted by the BH, carries into the BH not only its mass, but also its angular momentum $l_{\rm acc}$ and as a result the orbital inclination is altered. It is reasonable to assume that $l_{\rm acc}$ is going to be the angular momentum right at the innermost stable circular orbit (ISCO) of the BH, since inside the ISCO the infalling mass is following a radial plunge. The position of the ISCO depends on the BH mass and spin and on the infalling particle's orbital inclination. Therefore, during an accretion episode we need to find the position of the ISCO, calculate the angular momentum of the accreted mass at the ISCO, and then calculate the BH's new spin and mass and finally update the value of the inclination. 

The equation that describes the radial geodesic motion of a mass particle in the Kerr spacetime in terms of the Boyer-Lindquist coordinates $(t, r, \theta)$, is given by \cite{mtw}
\begin{equation} 
\Sigma^2 \left (\frac{dr}{d\tau} \right)^2=[E(r^2+a^2)-aL_z]^2-\Delta[r^2+(L_z-aE)^2+Q]\equiv R
\end{equation}

\begin{equation} 
\Sigma\left({dt\over d\tau}\right) = E\left[{(r^2+a^2)^2\over\Delta} - a^2\sin^2\theta\right] +aL_z\left(1 - {r^2+a^2\over\Delta}\right)
\end{equation} 

where $ \Sigma=r^2+a^2\cos^2\theta $ and $\Delta=r^2-2Mr+a^2$. The quantities $E$, $Lz$ and $Q$ (the three constants of motion) are the specific energy, angular momentum and Carter constant, with the Carter constant being defined as : $Q=\frac{L_z^2}{\cos^2i}-L_z^2=L_z^2\tan^2i$. Notice that all equations make use of geometrized units ($G=c=1$). A mass particle orbiting at the ISCO should fulfill three conditions:  $R=0$ (for the orbit to be circular), $R'=0$ (for the orbit to remain circular, i.e. zero radial acceleration) and finally $R''=0$ (for the orbit to be stable). By solving this set of equations ($R=R'=R''=0$) we get the position of the ISCO $r_{\rm ISCO}$, as well as the angular momentum ($L_{\rm ISCO}$) and energy ($E_{\rm ISCO}$) of the accreted mass. In our calculations the accretion of mass $m_{\rm acc}$ is treated as a series of accretion episodes of infinitesimal mass parcels $m_i$ ($\sum_i m_i=m_{\rm acc}$). After every accretion episode, the mass and angular momentum of the BH is updated and so is the inclination. We proceed to the next accretion episode using these updated values until $m_{\rm acc}$ has been accreted. 

\section{RESULTS}

We considered a large library of PS models (over 200), where we varied a number of PS model parameters (See Table~1).
For each of the models, we examined the PDF of post-SN BH spin-orbit misalignment angles in BH XRBs and its dependence on the masses of the BH and the donor star. We note here that in all the results we present here, we considered only the transient BH XRB population. This selection was made in order to facilitate a direct comparison with the currently observed sample of Galactic BH XRBs. The dynamical measurement of the BH mass, which is required in order to identify the compact object accretor of an XRB as BH, can only be done when the XRB is in quiescence. The same limitation holds for the two currently available observational technics (continuum-fitting and Fe K methods) for the measurement of the BH spin magnitude. 

\begin{figure}
\includegraphics[scale=.50]{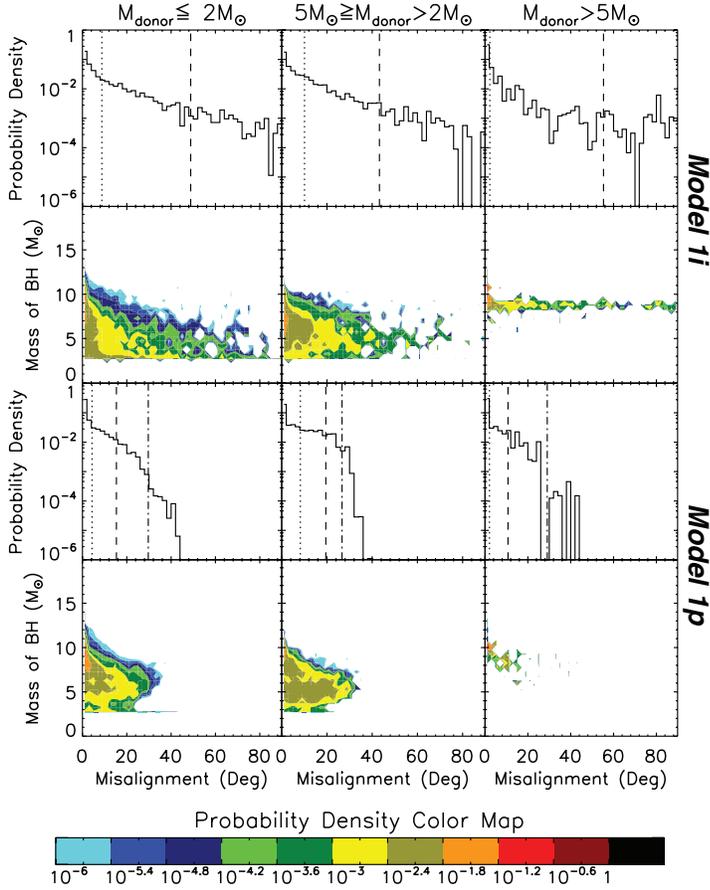}
\caption{PDF of the post-SN BH spin-orbit misalignment angle (first and third row of panels) and 2D PDF of the post-SN misalignment angle versus BH mass (second and fourth row of panels), split into three donor mass ranges (first column: $M_{\rm donor}\leq 2\rm\,M_{\odot}$, second column: $2\rm\,M_{\odot}<M_{\rm donor}\geq 5\rm\,M_{\odot}$, and third column: $M_{\rm donor}\geq 5\rm\,M_{\odot}$). The upper six panels correspond to our \emph{``standard''} model $1i$, while the the lower six panels correspond to model $1p$ which has the same PS parameters as the \emph{``standard''} model, except that the direction of the SN kicks is always aligned with the spin of the BH progenitor and not isotropic. The dotted vertical line shows the 67\% confidence level (i.e. 67\% of the systems have misalignment angle less than that indicate by the position of the line), while the dashed and dash-dotted indicate the 95.4\% and 99.7\% confidence level correspondingly. }
\end{figure}

In Figure~1 we present the PDF of the post-SN BH spin-orbit misalignment angle and the two-dimensional PDF of the post-SN misalignment angle versus BH mass, split into three donor mass ranges, for our \emph{``standard''} model $1i$, and for model $1p$. Model $1p$ has the same PS parameters as the \emph{``standard''} model, except that the direction of the SN kicks is always perpendicular to the orbital plane and not isotropic. One general characteristic of the PDFs is that they are heavily skewed towards small misalignment angles. This characteristic is the result of the combination of two factors. BHs are generally believed to receive asymmetric kicks of smaller magnitude compared to that of NS. On the other hand, large kick magnitudes, that would result in a large misalignment angle, usually lead to the disruption of the binary, and thus these systems are effectively filtered out of the population. The latter effect becomes more prominent in the case of polar kicks (model $1p$). One would initially expect that kicks always directed perpendicular to the orbital plane would result in larger misalignment angles. However, a large kick perpendicular to the orbital plane is also more efficient at disrupting the binary. Figure~1 shows that the binary disruption effect is dominant, and that only BHs that received a small kick manage to remain bound in a binary, and thus polar kicks lead to overall smaller misalignment angles.

The two-dimensional PDFs of the post-SN misalignment angle versus BH mass in Figure~1 show that massive BHs tend to have smaller misalignment kicks. This trend is common in all our models, and is a consequence of our prescription for the asymmetric kicks imparted to BHs. As described in Section~2.3, massive BHs receive on average smaller kicks, which in turn lead to smaller misalignment angles. Moreover, the higher the total mass of the binary is, the less the orbit will change from a given SN kick. The same physical mechanism is also associated with our finding that XRBs with massive donor stars have smaller BH spin tilts.

\begin{figure}
\includegraphics[scale=.50]{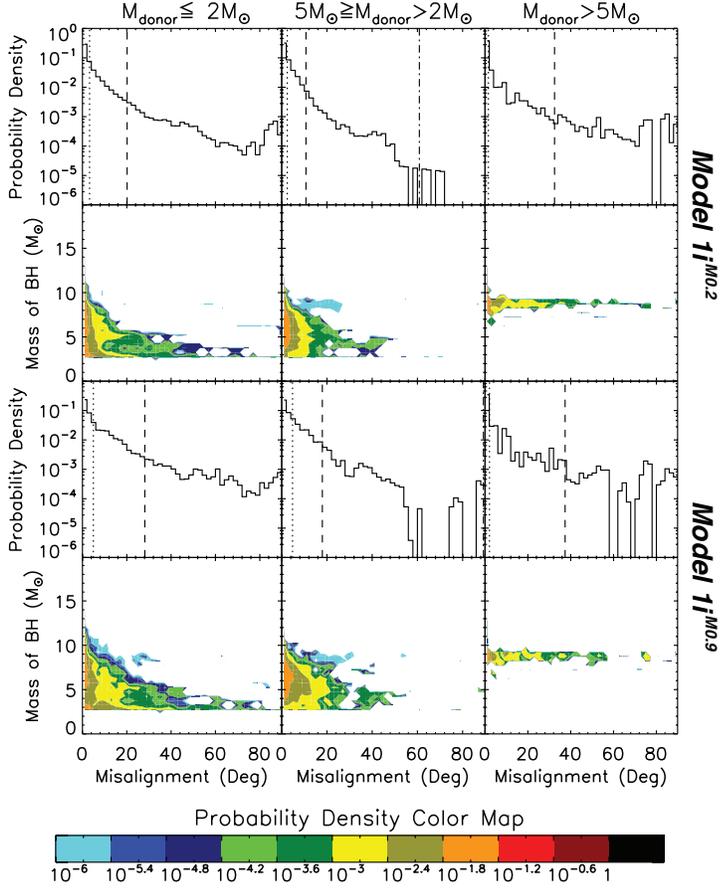}
\caption{Same as Figure~1, but taking into account the BH spin evolution due to accretion. The upper six panels correspond to model $1i^{M0.2}$, where the initial spin magnitude is drawn from a Maxwellian distribution with a maximum at 0.2 and truncated at 1. The lower six panels correspond to model $1i^{M0.9}$ which has the same PS parameters as model $1i^{M0.2}$, but the maximum of the Maxwellian initial BH spin magnitude distribution is at 0.9. }
\end{figure}

Figure~2 shows the PDF of the current BH spin-orbit misalignment angle and the two-dimensional PDF of the current misalignment angle versus BH mass, split into three donor mass ranges, for models $1i^{M0.2}$ and $1i^{M0.9}$. In this case, we take into account the evolution of the BH spin's direction and magnitude due to accretion during the XRB phase. The accretion of matter onto the BH always results on a gradual alignment of its spin with the orbital angular momentum. The efficiency of this process depends on the initial spin magnitude of the BH - the higher the initial BH spin is, the more difficult it is to change its direction via accretion. In Figure~2 we show two models with different distributions of initial BH spin magnitute, one with a Maxwellian distribution with a maximum at 0.2 and truncated at 1, and one where the maximum of the Maxwellian is at 0.9. Comparing these PDFs with the upper six panels of Figure~1, which show the PDFs of the post-SN misalignment angle of the same model, we see that for both models ($1i^{M0.2}$ and $1i^{M0.9}$) the PDFs of Figure~2 are shifted toward lower misalignment angles, and that indeed this effect is more prominent for smaller initial BH spin magnitudes (model $1i^{M0.2}$).

\begin{deluxetable}{lcccccccccccc}
\tablecolumns{13}
\tabletypesize{\scriptsize}
\tablecaption{67\%, 95.4\%, and 99.7\% confidence levels of BH spin-orbit misalignment angle PDFs of selected PS models. For each model, the confidence levels are reported for three donor mass ranges (column 2-4: $M_{\rm donor}\leq 2\rm\,M_{\odot}$, column 5-7: $2\rm\,M_{\odot}<M_{\rm donor}\geq 5\rm\,M_{\odot}$, and column 8-10: $M_{\rm donor}\geq 5\rm\,M_{\odot}$). Model names with no exponents correspond to the post-SN BH spin-orbit misalignment, right after the BH formation. For model names with exponents the BH spin evolution due to accretion was taken into account. For  the complete list of PS models see the online supplemental material.
\label{results}} 
\tablehead{    
   \colhead{} &
   \colhead{} &
   \multicolumn{3}{c}{$M_{\rm donor}\leq 2\rm\,M_{\odot}$} &
   \colhead{} &
   \multicolumn{3}{c}{$2\rm\,M_{\odot}<M_{\rm donor}<5\rm\,M_{\odot}$} &
   \colhead{} &
   \multicolumn{3}{c}{$M_{\rm donor}\geq 5\rm\,M_{\odot}$}  \\
   \cline{3-5} \cline{7-9} \cline{11-13} \\
     \colhead{PS Model} & 
     \colhead{} &
	 \colhead{67\% c.l.} &
	 \colhead{95.4\% c.l.} &
	 \colhead{99.7\% c.l.} &
     \colhead{} &
	 \colhead{67\% c.l.} &
	 \colhead{95.4\% c.l.} &
	 \colhead{99.7\% c.l.} &
     \colhead{} &
	 \colhead{67\% c.l.} &
	 \colhead{95.4\% c.l.} &
	 \colhead{99.7\% c.l.} 
	}
\startdata
$1i$ &  &   8.8 &  48.8 & 133.0 &  &  10.1 &  43.2 & 100.7 &  &   2.2 &  55.3 & 134.5 \\ 
$1p$ &  &   3.9 &  17.9 &  27.8 &  &  10.7 &  24.2 &  29.7 &  &   3.7 &  19.0 &  26.3 \\ 
$2i$ &  &  11.5 &  65.1 & 133.7 &  &   7.9 &  37.7 & 131.3 &  &   4.1 &  46.6 & 113.8 \\ 
$3i$ &  &  11.4 &  61.7 & 135.8 &  &  10.1 &  45.3 & 117.8 &  &   3.2 &  62.5 & 128.3 \\ 
$4i$ &  &   1.9 &  17.4 &  94.4 &  &   1.8 &   6.2 & 111.2 &  &   1.4 &  16.2 & 101.6 \\ 
$5i$ &  &   5.7 &  31.8 & 128.6 &  &   5.4 &  22.0 &  75.5 &  &   1.8 &  46.6 & 127.0 \\ 
$6i$ &  &  11.5 &  61.0 & 132.3 &  &  13.6 &  60.5 & 136.4 &  &   3.9 &  64.3 & 160.0 \\ 
$7i$ &  &  10.9 &  57.7 & 126.5 &  &  10.1 &  43.8 &  97.7 &  &   2.2 &  58.8 & 118.4 \\ 
$11i$ &  &   1.8 &  29.1 &  87.7 &  &   2.2 &  25.4 &  81.6 &  &   1.7 &   8.0 &  86.5 \\ 
$1i^{C0.2}$ &  &   3.2 &  19.0 &  90.0 &  &   2.3 &   9.1 &  69.8 &  &   1.7 &  28.8 & 123.6 \\ 
$1i^{C0.5}$ &  &   4.5 &  26.6 &  92.0 &  &   4.2 &  15.9 &  86.1 &  &   2.1 &  36.6 & 128.4 \\ 
$1i^{C0.9}$ &  &   5.7 &  32.7 &  92.6 &  &   6.0 &  21.7 &  89.3 &  &   2.1 &  39.5 & 133.7 \\ 
$1i^{M0.2}$ &  &   3.2 &  20.1 &  90.0 &  &   2.4 &  10.7 &  60.9 &  &   1.6 &  32.5 & 123.0 \\ 
$1i^{M0.5}$ &  &   4.3 &  26.4 &  92.5 &  &   4.3 &  17.3 &  86.1 &  &   2.0 &  36.7 & 127.5 \\ 
$1i^{M0.9}$ &  &   4.9 &  28.1 &  92.5 &  &   4.8 &  18.0 &  89.2 &  &   2.1 &  37.4 & 130.7 \\ 
$19i^{M0.5}$ &  &  15.3 &  52.2 & 118.4 &  &  11.6 &  50.8 &  93.4 &  &  12.8 &  41.2 & 118.0 \\ 

\enddata
\end{deluxetable}

Table~2 contains the 67\%, 95.4\%, and 99.7\% confidence levels of BH spin-orbit misalignment angle PDFs for a selected representative list of models (for  the complete list of PS models see the online supplemental material). After accounting for the partial alignment due to accretion onto the BH, we see that the majority of BH XRBs (at a 67\% confidence level) have BH spin-orbit misalignment angles below - and some times well below - $10^o$. This is a robust conclusion as it is true for \emph{all} our models. At the same time, we find that in most of our models there is a small part of the XRB population ($<5\%$) with misalignment angles above $20^o$. 

The main effect of varying the IMF or the stellar wind strength is on the BH mass spectrum. A flatter IMF or a weaker stellar wind lead to more massive BHs. This in turn, as we explained earlier, lead to smaller SN kicks and overall more massive binaries, and thus smaller BH spin-orbit misalignment angles. On the other hand, a variation of the CE efficiency does not seem to affect spin misalignment of our BH XRB population. Variations of the $\sigma$ parameter of the Maxwellian distribution for the SN kicks affect, as expected, the BH tilts, with higher kicks resulting to higher misalignment angles. However this effect saturates for very high kicks, as they lead to disruption rather than to very high misalignment angles.   

Finally, we examine the extreme case where BH receive the same kicks as NS, directly drawn  from a Maxwellian distribution and not scaled to the fall-back mass (Model 19). Although this model predicts unrealistically large BH kicks, it serves as an absolute upper limit on the possible BH spin-orbit misalignment. We find the even in this extreme case, the majority of the BH XRBs have misalignment angles below $\sim 15^o$.

\section{Summary and Conclusions}

We performed PS simulations of a Galactic field population of BH XRBs and studied the distribution of BH spin-orbit misalignment angles of transient, Roche-lobe overflowing BH XRBs. In the determination of the current misalignment angle we took into account the evolution of the BH spin magnitude and direction due to accretion of matter during the XRB phase. We examined over 200 PS models, varying PS parameters such as  IMF, stellar wind strength,  CE efficiency, SN kick's magnitude and direction distributions, and BH's initial spin magnitude distribution. We found that for the majority of BH XRBs in our model populations, the misalignment angle is below $10^o$, while at the same time there is a small part of the population ($<5\%$) that has misalignment angles higher than $20^o$ which in some cases can even exceed $90^o$. However this sub-population of highly misaligned BH XRBs account for less than $0.3\%$ of the total population. These results are robust among all our models, and all combinations of PS parameters. 

One caveat of our study is the assumption that the accretion disk around the BH is parallel to the orbital plane all the way down to the ISCO. \citet{NP1998} showed that the inner parts of an accretion disk around a spinning BH are forced to align with the spin of the BH by the Bardeen-Petterson effect, and that the same torque that aligns the inner disk with the BH tends also to align the spin of the BH with the outer accretion disk. Effectively this means that angular momentum is transfered to the BH from the outer parts of the disk. Although this mechanism depends on the details of the accretion disk model adopted \citep[see also][]{FA2005,King2005,LP2006, Fragile2007, MPT2007,Fragile2009}, the overall effect is to accelerate the alignment of the BH spin compared to our calculations. In this sense, our calculation of BH spin evolution due to accretion pose an upper limit on the current BH spin-orbit misalignment in Galactic BH XRBs.      

In recent years, several research groups have attempted to measure the BH spin magnitude in BH XRBs. Both currently available methods for the measurement of BH spins depend on the BH spin-orbit misalignment angle \citep{MR2009}. The X-ray continuum spectral fitting method \citep{Shafee2006} assumes a full alignment between the BH spin and the orbital angular momentum. The Fe K spectral line method \citep{BR2006} does not make any explicit assumption, but its accuracy could be improved from additional constraints on the BH spin tilts. This letter is the first study that provides theoretical constraints on the BH spin-orbit misalignment in BH XRBs.

\acknowledgements 
The authors would like to thank Prof. Kalogera, Dr. McClintock, and Dr. Mandel for their comments on the manuscript and other helpful discussions that greatly improved this work. TF acknowledges support from the Northwestern Presidential Fellowship.

\appendix

\begin{deluxetable}{lcccccccccccc}
\tablecolumns{13}
\tabletypesize{\scriptsize}
\tablecaption{Online only version of Table~2. 
\label{table}} 
\tablehead{    
   \colhead{} &
   \colhead{} &
   \multicolumn{3}{c}{$M_{\rm donor}\leq 2\rm\,M_{\odot}$} &
   \colhead{} &
   \multicolumn{3}{c}{$2\rm\,M_{\odot}<M_{\rm donor}<5\rm\,M_{\odot}$} &
   \colhead{} &
   \multicolumn{3}{c}{$M_{\rm donor}\geq 5\rm\,M_{\odot}$}  \\
   \cline{3-5} \cline{7-9} \cline{11-13} \\
     \colhead{PS Model} & 
     \colhead{} &
	 \colhead{67\% c.l.} &
	 \colhead{95.4\% c.l.} &
	 \colhead{99.7\% c.l.} &
     \colhead{} &
	 \colhead{67\% c.l.} &
	 \colhead{95.4\% c.l.} &
	 \colhead{99.7\% c.l.} &
     \colhead{} &
	 \colhead{67\% c.l.} &
	 \colhead{95.4\% c.l.} &
	 \colhead{99.7\% c.l.} 
	}
\startdata
$1i$ &  &   8.8 &  48.8 & 133.0 &  &  10.1 &  43.2 & 100.7 &  &   2.2 &  55.3 & 134.5 \\ 
$1i^{C0.2}$ &  &   3.2 &  19.0 &  90.0 &  &   2.3 &   9.1 &  69.8 &  &   1.7 &  28.8 & 123.6 \\ 
$1i^{C0.5}$ &  &   4.5 &  26.6 &  92.0 &  &   4.2 &  15.9 &  86.1 &  &   2.1 &  36.6 & 128.4 \\ 
$1i^{C0.9}$ &  &   5.7 &  32.7 &  92.6 &  &   6.0 &  21.7 &  89.3 &  &   2.1 &  39.5 & 133.7 \\ 
$1i^{M0.2}$ &  &   3.2 &  20.1 &  90.0 &  &   2.4 &  10.7 &  60.9 &  &   1.6 &  32.5 & 123.0 \\ 
$1i^{M0.5}$ &  &   4.3 &  26.4 &  92.5 &  &   4.3 &  17.3 &  86.1 &  &   2.0 &  36.7 & 127.5 \\ 
$1i^{M0.9}$ &  &   4.9 &  28.1 &  92.5 &  &   4.8 &  18.0 &  89.2 &  &   2.1 &  37.4 & 130.7 \\ 
$1p$ &  &   3.9 &  17.9 &  27.8 &  &  10.7 &  24.2 &  29.7 &  &   3.7 &  19.0 &  26.3 \\ 
$1p^{C0.2}$ &  &   2.1 &  11.3 &  20.6 &  &   4.1 &  13.5 &  22.8 &  &   1.4 &   8.0 &  13.7 \\ 
$1p^{C0.5}$ &  &   2.7 &  13.9 &  23.4 &  &   6.4 &  16.9 &  24.9 &  &   2.7 &  12.5 &  20.5 \\ 
$1p^{C0.9}$ &  &   3.1 &  15.5 &  24.9 &  &   8.2 &  19.5 &  26.6 &  &   4.3 &  15.5 &  22.9 \\ 
$1p^{M0.2}$ &  &   2.1 &  11.4 &  21.6 &  &   4.1 &  13.6 &  22.7 &  &   1.6 &   8.4 &  15.5 \\ 
$1p^{M0.5}$ &  &   2.6 &  13.9 &  24.3 &  &   6.2 &  17.3 &  25.4 &  &   2.8 &  12.8 &  20.2 \\ 
$1p^{M0.9}$ &  &   2.8 &  14.6 &  24.7 &  &   7.1 &  18.1 &  25.2 &  &   2.9 &  13.6 &  22.8 \\ 
$2i$ &  &  11.5 &  65.1 & 133.7 &  &   7.9 &  37.7 & 131.3 &  &   4.1 &  46.6 & 113.8 \\ 
$2i^{C0.2}$ &  &   4.4 &  33.7 & 119.7 &  &   2.0 &   8.5 & 103.7 &  &   1.9 &  33.5 & 115.1 \\ 
$2i^{C0.5}$ &  &   7.2 &  52.7 & 126.6 &  &   3.5 &  14.2 & 118.3 &  &   3.5 &  45.8 & 118.6 \\ 
$2i^{C0.9}$ &  &   8.9 &  58.6 & 128.8 &  &   4.7 &  20.5 & 123.7 &  &   3.9 &  48.6 & 120.1 \\ 
$2i^{M0.2}$ &  &   4.4 &  33.7 & 121.8 &  &   2.0 &  10.2 & 111.2 &  &   2.1 &  34.5 & 100.3 \\ 
$2i^{M0.5}$ &  &   6.9 &  50.3 & 128.8 &  &   3.4 &  15.1 & 119.6 &  &   3.2 &  45.6 & 118.5 \\ 
$2i^{M0.9}$ &  &   8.1 &  57.7 & 115.3 &  &   4.0 &  16.7 & 119.2 &  &   3.5 &  48.0 & 118.7 \\ 
$3i$ &  &  11.4 &  61.7 & 135.8 &  &  10.1 &  45.3 & 117.8 &  &   3.2 &  62.5 & 128.3 \\ 
$3i^{C0.2}$ &  &   3.9 &  27.9 & 103.6 &  &   2.7 &  11.1 &  90.7 &  &   1.9 &  89.7 & 124.7 \\ 
$3i^{C0.5}$ &  &   6.1 &  38.8 & 111.8 &  &   4.9 &  17.5 &  91.7 &  &   2.2 &  89.8 & 125.2 \\ 
$3i^{C0.9}$ &  &   7.8 &  44.6 & 119.1 &  &   7.1 &  24.3 &  94.5 &  &   3.5 &  89.8 & 125.2 \\ 
$3i^{M0.2}$ &  &   4.0 &  28.0 & 107.1 &  &   2.7 &  11.2 &  90.3 &  &   1.9 &  87.5 & 124.2 \\ 
$3i^{M0.5}$ &  &   6.0 &  35.9 & 111.6 &  &   4.8 &  18.3 &  91.0 &  &   2.0 &  89.8 & 125.2 \\ 
$3i^{M0.9}$ &  &   6.7 &  41.0 & 112.9 &  &   5.6 &  21.1 &  93.3 &  &   3.1 &  89.8 & 125.2 \\ 
$4i$ &  &   1.9 &  17.4 &  94.4 &  &   1.8 &   6.2 & 111.2 &  &   1.4 &  16.2 & 101.6 \\ 
$4i^{C0.2}$ &  &   1.7 &   8.7 &  51.1 &  &   1.5 &   2.8 &  90.5 &  &   1.3 &  19.9 &  94.8 \\ 
$4i^{C0.5}$ &  &   1.8 &  12.4 &  53.2 &  &   1.6 &   4.0 &  90.5 &  &   1.4 &  24.0 &  98.7 \\ 
$4i^{C0.9}$ &  &   1.9 &  14.6 &  55.7 &  &   1.7 &   4.5 &  93.0 &  &   1.4 &  25.6 &  99.3 \\ 
$4i^{M0.2}$ &  &   1.7 &   9.0 &  51.6 &  &   1.5 &   3.1 &  90.5 &  &   1.3 &  19.6 &  96.6 \\ 
$4i^{M0.5}$ &  &   1.8 &  12.8 &  53.9 &  &   1.6 &   4.1 &  91.2 &  &   1.3 &  21.8 &  97.6 \\ 
$4i^{M0.9}$ &  &   1.8 &  13.6 &  55.3 &  &   1.6 &   4.3 &  90.7 &  &   1.3 &  25.5 &  99.3 \\ 
$4p$ &  &   2.8 &  11.3 &  24.6 &  &   3.6 &   8.9 &  13.5 &  &   1.1 &   3.5 &   6.7 \\ 
$4p^{C0.2}$ &  &   1.8 &   6.5 &  14.5 &  &   1.6 &   4.0 &   8.0 &  &   1.1 &   2.2 &   4.4 \\ 
$4p^{C0.5}$ &  &   2.1 &   8.3 &  18.8 &  &   2.1 &   5.3 &   9.2 &  &   1.1 &   2.2 &   6.6 \\ 
$4p^{C0.9}$ &  &   2.4 &   9.5 &  21.8 &  &   2.7 &   6.5 &  10.9 &  &   1.1 &   2.9 &   6.7 \\ 
$4p^{M0.2}$ &  &   1.7 &   6.6 &  14.9 &  &   1.7 &   4.1 &   8.2 &  &   1.1 &   2.2 &   4.4 \\ 
$4p^{M0.5}$ &  &   2.0 &   8.4 &  18.1 &  &   2.1 &   5.8 &  10.0 &  &   1.1 &   2.6 &   4.9 \\ 
$4p^{M0.9}$ &  &   2.2 &   8.7 &  21.8 &  &   2.2 &   5.9 &  10.6 &  &   1.1 &   2.2 &   6.7 \\ 
$5i$ &  &   5.7 &  31.8 & 128.6 &  &   5.4 &  22.0 &  75.5 &  &   1.8 &  46.6 & 127.0 \\ 
$5i^{C0.2}$ &  &   2.5 &  16.0 &  91.9 &  &   1.8 &   6.7 &  17.4 &  &   1.3 &  33.2 & 113.7 \\ 
$5i^{C0.5}$ &  &   3.8 &  21.8 &  97.1 &  &   2.5 &  10.5 &  22.5 &  &   1.4 &  40.2 & 121.4 \\ 
$5i^{C0.9}$ &  &   4.4 &  25.6 & 102.6 &  &   3.6 &  13.8 &  33.4 &  &   1.6 &  43.0 & 123.9 \\ 
$5i^{M0.2}$ &  &   2.5 &  16.8 &  91.2 &  &   1.9 &   7.0 &  17.9 &  &   1.3 &  32.8 & 111.4 \\ 
$5i^{M0.5}$ &  &   3.7 &  21.8 &  97.3 &  &   2.5 &  10.2 &  24.9 &  &   1.4 &  37.8 & 122.9 \\ 
$5i^{M0.9}$ &  &   4.0 &  23.7 & 103.7 &  &   3.0 &  12.3 &  30.8 &  &   1.4 &  40.6 & 123.1 \\ 
$5p$ &  &   4.3 &  15.3 &  29.7 &  &   8.2 &  19.5 &  26.7 &  &   2.1 &  10.7 &  29.0 \\ 
$5p^{C0.2}$ &  &   2.4 &   9.7 &  23.6 &  &   2.9 &   8.8 &  16.4 &  &   1.0 &   4.6 &  22.7 \\ 
$5p^{C0.5}$ &  &   3.3 &  11.4 &  26.9 &  &   4.7 &  11.8 &  18.9 &  &   1.2 &   6.2 &  27.1 \\ 
$5p^{C0.9}$ &  &   3.9 &  12.9 &  28.8 &  &   6.2 &  14.0 &  20.6 &  &   1.3 &   6.7 &  28.9 \\ 
$5p^{M0.2}$ &  &   2.4 &   9.9 &  23.8 &  &   2.7 &   9.1 &  17.1 &  &   0.9 &   5.5 &  23.8 \\ 
$5p^{M0.5}$ &  &   3.3 &  11.7 &  26.8 &  &   4.5 &  12.4 &  19.2 &  &   1.3 &   5.6 &  27.7 \\ 
$5p^{M0.9}$ &  &   3.5 &  11.9 &  27.4 &  &   5.4 &  13.0 &  20.2 &  &   1.3 &   6.2 &  29.0 \\ 
$6i$ &  &  11.5 &  61.0 & 132.3 &  &  13.6 &  60.5 & 136.4 &  &   3.9 &  64.3 & 160.0 \\ 
$6i^{C0.2}$ &  &   4.5 &  31.8 & 102.7 &  &   3.5 &  17.7 &  91.3 &  &   1.9 &  21.5 & 110.1 \\ 
$6i^{C0.5}$ &  &   7.2 &  47.1 & 109.7 &  &   6.1 &  28.5 &  92.4 &  &   1.9 &  29.6 & 117.1 \\ 
$6i^{C0.9}$ &  &   9.0 &  52.7 & 113.8 &  &   8.5 &  38.4 &  95.4 &  &   3.0 &  34.1 & 117.1 \\ 
$6i^{M0.2}$ &  &   4.5 &  36.4 & 101.1 &  &   3.6 &  18.7 &  91.3 &  &   1.9 &  21.9 & 113.3 \\ 
$6i^{M0.5}$ &  &   6.7 &  45.9 & 109.7 &  &   5.8 &  31.0 &  91.4 &  &   1.9 &  24.7 & 117.1 \\ 
$6i^{M0.9}$ &  &   7.9 &  51.7 & 112.8 &  &   6.8 &  34.6 &  92.6 &  &   2.1 &  33.9 & 117.1 \\ 
$6p$ &  &   4.0 &  20.5 &  30.3 &  &  10.4 &  26.3 &  32.0 &  &   4.0 &  22.9 &  35.4 \\ 
$6p^{C0.2}$ &  &   2.2 &  13.4 &  26.0 &  &   4.4 &  14.7 &  25.7 &  &   1.5 &  10.8 &  22.6 \\ 
$6p^{C0.5}$ &  &   2.8 &  15.8 &  28.8 &  &   6.7 &  18.9 &  27.3 &  &   2.2 &  19.0 &  22.8 \\ 
$6p^{C0.9}$ &  &   3.1 &  17.5 &  30.3 &  &   8.1 &  21.7 &  28.9 &  &   2.4 &  21.1 &  27.2 \\ 
$6p^{M0.2}$ &  &   3.6 &  12.8 &  23.1 &  &   4.3 &  15.1 &  25.9 &  &   1.5 &   9.5 &  22.6 \\ 
$6p^{M0.5}$ &  &   5.5 &  15.8 &  25.8 &  &   6.1 &  19.2 &  27.6 &  &   2.2 &  18.6 &  23.5 \\ 
$6p^{M0.9}$ &  &   7.1 &  17.3 &  26.0 &  &   7.1 &  20.5 &  28.4 &  &   2.0 &  20.6 &  22.9 \\ 
$7i$ &  &  10.9 &  57.7 & 126.5 &  &  10.1 &  43.8 &  97.7 &  &   2.2 &  58.8 & 118.4 \\ 
$7i^{C0.2}$ &  &   3.9 &  29.6 &  97.6 &  &   2.3 &  10.8 &  91.2 &  &   1.4 &  45.2 & 105.7 \\ 
$7i^{C0.5}$ &  &   6.0 &  42.4 & 108.5 &  &   4.1 &  19.0 &  91.3 &  &   1.6 &  53.2 & 107.5 \\ 
$7i^{C0.9}$ &  &   7.7 &  51.0 & 128.9 &  &   5.7 &  25.2 &  92.2 &  &   1.9 &  55.6 & 108.4 \\ 
$7i^{M0.2}$ &  &   3.9 &  28.6 &  98.9 &  &   2.2 &  12.9 &  91.1 &  &   1.6 &  46.4 & 106.6 \\ 
$7i^{M0.5}$ &  &   6.0 &  42.4 & 107.9 &  &   4.1 &  17.2 &  91.3 &  &   1.7 &  52.5 & 107.7 \\ 
$7i^{M0.9}$ &  &   6.9 &  48.1 & 122.2 &  &   4.5 &  21.3 &  92.4 &  &   1.9 &  55.2 & 108.4 \\ 
$7p$ &  &   5.7 &  19.6 &  29.4 &  &  11.5 &  24.0 &  30.0 &  &   7.5 &  20.4 &  26.7 \\ 
$7p^{C0.2}$ &  &   3.3 &  13.6 &  24.4 &  &   4.2 &  12.8 &  22.8 &  &   2.1 &   9.8 &  18.3 \\ 
$7p^{C0.5}$ &  &   4.3 &  16.3 &  26.8 &  &   6.9 &  16.8 &  24.9 &  &   3.8 &  12.1 &  24.2 \\ 
$7p^{C0.9}$ &  &   4.8 &  17.8 &  27.6 &  &   9.0 &  20.0 &  26.4 &  &   4.0 &  15.2 &  26.4 \\ 
$7p^{M0.2}$ &  &   3.2 &  13.9 &  24.3 &  &   4.3 &  14.5 &  24.1 &  &   1.9 &   9.8 &  20.1 \\ 
$7p^{M0.5}$ &  &   4.2 &  15.9 &  26.1 &  &   6.7 &  17.5 &  25.5 &  &   3.4 &  15.2 &  22.4 \\ 
$7p^{M0.9}$ &  &   4.5 &  17.5 &  27.3 &  &   7.6 &  18.0 &  25.1 &  &   4.0 &  14.5 &  21.6 \\ 
$8i$ &  &   2.2 &  25.2 &  78.6 &  &   1.9 &   6.9 &  83.1 &  &   1.7 &  26.0 & 112.9 \\ 
$8i^{C0.2}$ &  &   1.8 &  16.4 &  34.1 &  &   1.5 &   3.3 &  28.8 &  &   1.5 &  16.1 & 112.4 \\ 
$8i^{C0.5}$ &  &   2.0 &  24.0 &  42.0 &  &   1.7 &   4.3 &  68.5 &  &   1.5 &  18.2 & 113.1 \\ 
$8i^{C0.9}$ &  &   2.1 &  25.0 &  67.0 &  &   1.8 &   5.3 &  78.3 &  &   1.5 &  20.5 & 113.1 \\ 
$8i^{M0.2}$ &  &   1.8 &  18.4 &  34.4 &  &   1.5 &   3.5 &  50.0 &  &   1.5 &  16.6 & 112.4 \\ 
$8i^{M0.5}$ &  &   2.0 &  23.1 &  38.4 &  &   1.7 &   4.4 &  61.5 &  &   1.5 &  18.8 & 113.1 \\ 
$8i^{M0.9}$ &  &   2.0 &  24.2 &  68.0 &  &   1.7 &   4.5 &  78.6 &  &   1.5 &  19.9 & 113.1 \\ 
$8p$ &  &   3.3 &  13.6 &  25.1 &  &   3.9 &   9.5 &  15.5 &  &   1.4 &   6.0 &  13.7 \\ 
$8p^{C0.2}$ &  &   1.8 &   7.5 &  14.9 &  &   1.7 &   4.4 &   8.4 &  &   0.9 &   2.1 &   9.4 \\ 
$8p^{C0.5}$ &  &   2.2 &   9.9 &  17.9 &  &   2.2 &   6.2 &  10.1 &  &   1.0 &   2.2 &  13.6 \\ 
$8p^{C0.9}$ &  &   2.6 &  11.5 &  20.1 &  &   3.1 &   7.2 &  11.6 &  &   1.0 &   3.0 &  15.9 \\ 
$8p^{M0.2}$ &  &   1.9 &   7.4 &  15.4 &  &   1.7 &   4.7 &   8.7 &  &   0.9 &   2.1 &  10.2 \\ 
$8p^{M0.5}$ &  &   2.1 &  10.2 &  19.0 &  &   2.2 &   6.3 &   9.8 &  &   1.0 &   2.2 &  12.8 \\ 
$8p^{M0.9}$ &  &   2.2 &  10.9 &  20.1 &  &   2.7 &   6.5 &  10.6 &  &   1.0 &   3.0 &  15.9 \\ 
$9i$ &  &   7.3 &  29.7 & 103.7 &  &   6.3 &  23.4 & 102.3 &  &   2.2 &  37.0 & 144.5 \\ 
$9i^{C0.2}$ &  &   2.7 &  16.0 &  94.7 &  &   2.0 &   7.6 &  23.2 &  &   2.1 &  28.1 & 123.6 \\ 
$9i^{C0.5}$ &  &   4.2 &  20.9 &  99.3 &  &   3.2 &  12.2 &  89.6 &  &   3.2 &  31.7 & 137.3 \\ 
$9i^{C0.9}$ &  &   5.2 &  24.4 & 101.6 &  &   4.3 &  17.0 &  89.6 &  &   3.6 &  34.3 & 140.4 \\ 
$9i^{M0.2}$ &  &   2.8 &  16.7 &  96.8 &  &   2.1 &   8.5 &  89.6 &  &   1.9 &  29.2 & 125.9 \\ 
$9i^{M0.5}$ &  &   4.0 &  21.7 & 101.6 &  &   3.2 &  11.8 &  89.6 &  &   2.9 &  29.7 & 139.6 \\ 
$9i^{M0.9}$ &  &   4.5 &  22.8 & 101.6 &  &   3.7 &  13.9 &  89.6 &  &   3.3 &  33.3 & 138.3 \\ 
$9p$ &  &   5.1 &  17.0 &  34.1 &  &   9.0 &  19.5 &  26.3 &  &   2.7 &  13.0 &  20.5 \\ 
$9p^{C0.2}$ &  &   3.1 &  10.6 &  25.6 &  &   3.4 &   9.0 &  15.8 &  &   0.3 &   4.4 &  15.0 \\ 
$9p^{C0.5}$ &  &   4.2 &  12.9 &  30.6 &  &   5.4 &  12.1 &  17.9 &  &   0.5 &   4.5 &  23.5 \\ 
$9p^{C0.9}$ &  &   5.0 &  14.2 &  33.3 &  &   6.9 &  14.3 &  20.0 &  &   0.7 &   6.8 &  27.2 \\ 
$9p^{M0.2}$ &  &   3.1 &  10.8 &  26.3 &  &   3.5 &   9.6 &  16.1 &  &   0.3 &   4.4 &  12.2 \\ 
$9p^{M0.5}$ &  &   4.2 &  13.0 &  30.6 &  &   5.4 &  12.3 &  18.2 &  &   0.5 &   4.5 &  23.1 \\ 
$9p^{M0.9}$ &  &   4.5 &  13.6 &  32.3 &  &   6.0 &  13.4 &  20.2 &  &   0.7 &   6.7 &  26.1 \\
$10i$ &  &  13.5 &  71.6 & 139.4 &  &  13.3 &  64.0 & 138.2 &  &   3.5 &  39.1 & 133.3 \\ 
$10i^{C0.2}$ &  &   5.5 &  38.8 &  98.3 &  &   3.7 &  20.6 & 104.5 &  &   1.9 &  27.8 & 104.2 \\ 
$10i^{C0.5}$ &  &   8.7 &  50.4 & 108.3 &  &   6.7 &  30.8 & 121.9 &  &   2.3 &  34.0 & 112.3 \\ 
$10i^{C0.9}$ &  &  11.0 &  55.5 & 114.9 &  &   9.5 &  43.2 & 129.7 &  &   2.5 &  37.5 & 117.9 \\ 
$10i^{M0.2}$ &  &   5.5 &  38.5 &  97.3 &  &   3.8 &  21.5 & 116.4 &  &   1.9 &  31.5 & 106.2 \\ 
$10i^{M0.5}$ &  &   8.4 &  50.8 & 105.8 &  &   6.4 &  31.1 & 124.9 &  &   2.3 &  35.1 & 108.3 \\ 
$10i^{M0.9}$ &  &   9.6 &  54.2 & 108.7 &  &   7.6 &  34.4 & 119.9 &  &   3.2 &  36.6 & 111.3 \\ 
$10p$ &  &   4.5 &  20.0 &  33.1 &  &  12.3 &  26.4 &  33.0 &  &   8.2 &  26.5 &  29.7 \\ 
$10p^{C0.2}$ &  &   3.7 &  14.3 &  25.2 &  &   5.4 &  15.6 &  23.8 &  &   2.8 &  11.9 &  16.1 \\ 
$10p^{C0.5}$ &  &   4.8 &  17.1 &  28.3 &  &   8.7 &  19.1 &  25.2 &  &   5.2 &  16.5 &  25.2 \\ 
$10p^{C0.9}$ &  &   5.5 &  18.8 &  29.5 &  &  11.0 &  21.8 &  27.4 &  &   7.4 &  17.8 &  27.7 \\ 
$10p^{M0.2}$ &  &   3.6 &  14.5 &  25.2 &  &   4.8 &  15.0 &  22.9 &  &   3.3 &  14.2 &  16.0 \\ 
$10p^{M0.5}$ &  &   4.6 &  17.0 &  29.0 &  &   8.4 &  19.4 &  25.8 &  &   3.7 &  17.2 &  25.4 \\ 
$10p^{M0.9}$ &  &   5.2 &  18.0 &  29.3 &  &   9.4 &  20.7 &  26.9 &  &   6.0 &  17.8 &  27.5 \\ 
$11i$ &  &   1.8 &  29.1 &  87.7 &  &   2.2 &  25.4 &  81.6 &  &   1.7 &   8.0 &  86.5 \\ 
$11i^{C0.2}$ &  &   1.3 &  13.3 &  83.3 &  &   1.1 &   6.4 &  56.5 &  &   1.5 &   2.1 &  46.2 \\ 
$11i^{C0.5}$ &  &   1.6 &  18.5 &  85.8 &  &   1.5 &  10.5 &  76.0 &  &   1.5 &   2.7 &  54.4 \\ 
$11i^{C0.9}$ &  &   1.7 &  22.1 &  87.0 &  &   1.8 &  14.2 &  78.1 &  &   1.5 &   4.0 &  55.9 \\ 
$11i^{M0.2}$ &  &   1.3 &  13.4 &  82.1 &  &   1.2 &   6.7 &  58.5 &  &   1.5 &   2.2 &  48.0 \\ 
$11i^{M0.5}$ &  &   1.6 &  18.5 &  85.4 &  &   1.4 &  11.0 &  60.7 &  &   1.5 &   2.7 &  49.2 \\ 
$11i^{M0.9}$ &  &   1.6 &  19.8 &  86.8 &  &   1.6 &  12.3 &  58.6 &  &   1.5 &   3.5 &  55.6 \\ 
$11p$ &  &   0.4 &  14.1 &  25.0 &  &   1.1 &  19.6 &  28.6 &  &   0.0 &   8.6 &  27.8 \\ 
$11p^{C0.2}$ &  &   0.2 &   8.7 &  20.8 &  &   0.6 &   8.1 &  17.4 &  &   0.0 &   2.2 &   6.9 \\ 
$11p^{C0.5}$ &  &   0.2 &  10.6 &  23.0 &  &   0.9 &  11.4 &  20.0 &  &   0.0 &   4.2 &  13.1 \\ 
$11p^{C0.9}$ &  &   0.2 &  11.9 &  24.3 &  &   1.1 &  14.2 &  22.2 &  &   0.0 &   6.4 &  17.5 \\ 
$11p^{M0.2}$ &  &   0.2 &   8.8 &  21.6 &  &   0.6 &   8.4 &  17.5 &  &   0.0 &   2.2 &  10.2 \\ 
$11p^{M0.5}$ &  &   0.2 &  10.7 &  23.2 &  &   0.8 &  11.5 &  20.3 &  &   0.0 &   4.4 &  15.2 \\ 
$11p^{M0.9}$ &  &   0.2 &  11.0 &  23.4 &  &   0.9 &  12.9 &  21.6 &  &   0.0 &   5.5 &  19.5 \\ 
$12i$ &  &   0.7 &   5.6 &  53.0 &  &   0.6 &   3.9 &   9.1 &  &   1.7 &   1.9 &  35.3 \\ 
$12i^{C0.2}$ &  &   0.6 &   2.8 &  24.5 &  &   0.6 &   2.1 &   4.5 &  &   0.0 &   2.0 &  26.8 \\ 
$12i^{C0.5}$ &  &   0.7 &   3.9 &  38.0 &  &   0.6 &   2.3 &   6.4 &  &   0.0 &   2.0 &  31.7 \\ 
$12i^{C0.9}$ &  &   0.7 &   4.4 &  43.6 &  &   0.6 &   3.0 &   7.9 &  &   0.0 &   2.0 &  34.6 \\ 
$12i^{M0.2}$ &  &   0.6 &   2.9 &  21.9 &  &   0.6 &   2.1 &   5.1 &  &   1.5 &   2.0 &  27.8 \\ 
$12i^{M0.5}$ &  &   0.7 &   3.9 &  35.7 &  &   0.6 &   2.3 &   6.8 &  &   1.5 &   2.0 &  31.2 \\ 
$12i^{M0.9}$ &  &   0.7 &   4.1 &  41.4 &  &   0.6 &   2.4 &   7.0 &  &   1.5 &   2.0 &  32.8 \\ 
$12p$ &  &   0.9 &   7.6 &  23.5 &  &   1.1 &   6.8 &  12.8 &  &   0.0 &   1.9 &   4.6 \\ 
$12p^{C0.2}$ &  &   0.7 &   4.2 &  12.8 &  &   0.6 &   2.8 &   6.5 &  &   0.0 &   1.7 &   2.3 \\ 
$12p^{C0.5}$ &  &   0.8 &   5.3 &  17.6 &  &   0.7 &   4.1 &   7.9 &  &   0.0 &   1.7 &   3.0 \\ 
$12p^{C0.9}$ &  &   0.9 &   6.2 &  20.7 &  &   0.9 &   5.1 &   9.0 &  &   0.0 &   1.9 &   4.3 \\ 
$12p^{M0.2}$ &  &   0.7 &   4.2 &  13.6 &  &   0.6 &   3.0 &   6.6 &  &   0.0 &   1.7 &   2.3 \\ 
$12p^{M0.5}$ &  &   0.8 &   5.4 &  16.6 &  &   0.7 &   4.2 &   8.2 &  &   0.0 &   1.8 &   4.0 \\ 
$12p^{M0.9}$ &  &   0.8 &   5.7 &  19.5 &  &   0.8 &   4.5 &   8.6 &  &   0.0 &   1.9 &   4.2 \\ 
$13i$ &  &   1.5 &  18.3 &  69.1 &  &   1.6 &  14.9 &  40.5 &  &   1.7 &   4.5 &  62.2 \\ 
$13i^{C0.2}$ &  &   1.2 &   8.3 &  43.3 &  &   0.9 &   4.2 &  15.2 &  &   1.5 &   2.0 &  43.3 \\ 
$13i^{C0.5}$ &  &   1.4 &  11.7 &  56.9 &  &   1.1 &   6.8 &  21.8 &  &   1.5 &   2.2 &  50.5 \\ 
$13i^{C0.9}$ &  &   1.5 &  13.9 &  62.7 &  &   1.3 &   9.3 &  28.0 &  &   1.5 &   3.0 &  53.4 \\ 
$13i^{M0.2}$ &  &   1.2 &   8.7 &  45.7 &  &   0.9 &   4.4 &  15.8 &  &   1.5 &   2.1 &  40.6 \\ 
$13i^{M0.5}$ &  &   1.4 &  12.0 &  60.8 &  &   1.1 &   6.8 &  22.2 &  &   1.5 &   2.2 &  51.8 \\ 
$13i^{M0.9}$ &  &   1.4 &  12.8 &  58.3 &  &   1.1 &   8.2 &  25.2 &  &   1.5 &   2.3 &  51.4 \\ 
$13p$ &  &   0.8 &  11.0 &  21.6 &  &   1.5 &  15.8 &  24.9 &  &   0.0 &   5.8 &  20.9 \\ 
$13p^{C0.2}$ &  &   0.6 &   6.7 &  14.7 &  &   0.7 &   6.3 &  14.5 &  &   0.0 &   2.0 &  10.7 \\ 
$13p^{C0.5}$ &  &   0.7 &   8.4 &  17.0 &  &   1.1 &   9.0 &  17.1 &  &   0.0 &   3.0 &  14.0 \\ 
$13p^{C0.9}$ &  &   0.8 &   9.4 &  18.1 &  &   1.3 &  11.3 &  18.6 &  &   0.0 &   4.3 &  16.4 \\ 
$13p^{M0.2}$ &  &   0.6 &   6.7 &  15.4 &  &   0.7 &   6.5 &  14.8 &  &   0.0 &   1.9 &  10.7 \\ 
$13p^{M0.5}$ &  &   0.7 &   8.4 &  16.7 &  &   1.0 &   9.3 &  17.2 &  &   0.0 &   3.1 &  16.0 \\ 
$13p^{M0.9}$ &  &   0.8 &   8.8 &  17.8 &  &   1.2 &  10.2 &  18.0 &  &   0.0 &   3.5 &  16.2 \\ 
$14i$ &  &   1.1 &  34.0 & 112.6 &  &   2.6 &  36.0 & 100.1 &  &   1.7 &   9.4 & 102.1 \\ 
$14i^{C0.2}$ &  &   0.6 &  12.5 &  92.8 &  &   1.2 &   8.9 &  90.1 &  &   0.0 &   4.0 &  43.2 \\ 
$14i^{C0.5}$ &  &   0.8 &  19.9 & 100.6 &  &   1.6 &  15.8 &  90.5 &  &   0.0 &   6.2 &  55.4 \\ 
$14i^{C0.9}$ &  &   0.9 &  25.7 & 106.0 &  &   1.9 &  21.8 &  90.9 &  &   0.0 &   8.4 &  59.6 \\ 
$14i^{M0.2}$ &  &   0.6 &  13.7 &  93.2 &  &   1.2 &   9.7 &  89.9 &  &   1.5 &   3.8 &  48.3 \\ 
$14i^{M0.5}$ &  &   0.8 &  19.9 & 104.3 &  &   1.5 &  16.2 &  90.5 &  &   1.5 &   6.3 &  59.6 \\ 
$14i^{M0.9}$ &  &   0.8 &  22.1 & 104.3 &  &   1.7 &  19.2 &  90.6 &  &   1.5 &   9.5 &  59.6 \\ 
$14p$ &  &   0.4 &  15.9 &  28.5 &  &   1.1 &  21.9 &  31.5 &  &   0.0 &  13.3 &  33.1 \\ 
$14p^{C0.2}$ &  &   0.2 &  10.2 &  23.9 &  &   1.1 &  10.2 &  23.5 &  &   0.0 &   3.9 &   9.8 \\ 
$14p^{C0.5}$ &  &   0.2 &  12.3 &  26.3 &  &   1.1 &  14.2 &  26.6 &  &   0.0 &   6.7 &  18.8 \\ 
$14p^{C0.9}$ &  &   0.2 &  13.7 &  27.4 &  &   1.1 &  17.1 &  28.9 &  &   0.0 &  10.5 &  23.6 \\ 
$14p^{M0.2}$ &  &   0.2 &  10.3 &  24.2 &  &   0.9 &  10.1 &  24.0 &  &   0.0 &   4.0 &  11.0 \\ 
$14p^{M0.5}$ &  &   0.2 &  12.4 &  26.5 &  &   0.9 &  14.8 &  27.7 &  &   0.0 &   6.7 &  16.8 \\ 
$14p^{M0.9}$ &  &   0.2 &  12.9 &  26.9 &  &   0.9 &  15.6 &  27.8 &  &   0.0 &   9.8 &  23.2 \\ 
$15i$ &  &   2.9 &  35.2 & 115.2 &  &   2.9 &  29.4 &  96.9 &  &   1.7 &   8.2 &  83.9 \\ 
$15i^{C0.2}$ &  &   1.6 &  14.0 &  89.6 &  &   1.3 &   7.1 &  79.9 &  &   0.0 &   3.3 &  89.8 \\ 
$15i^{C0.5}$ &  &   2.0 &  20.4 &  90.9 &  &   1.8 &  12.6 &  89.9 &  &   0.0 &   5.4 &  92.1 \\ 
$15i^{C0.9}$ &  &   2.2 &  25.0 &  94.3 &  &   2.1 &  18.0 &  90.1 &  &   0.0 &   7.5 &  92.6 \\ 
$15i^{M0.2}$ &  &   1.6 &  14.5 &  89.5 &  &   1.4 &   7.8 &  86.4 &  &   1.5 &   3.1 &  90.2 \\ 
$15i^{M0.5}$ &  &   1.9 &  20.2 &  90.3 &  &   1.7 &  12.2 &  89.9 &  &   1.5 &   6.0 &  91.8 \\ 
$15i^{M0.9}$ &  &   2.1 &  23.2 &  92.1 &  &   1.9 &  15.0 &  89.9 &  &   1.5 &   5.4 &  92.9 \\ 
$15p$ &  &   0.9 &  14.1 &  26.0 &  &   1.6 &  20.4 &  28.8 &  &   0.0 &   5.9 &  19.8 \\ 
$15p^{C0.2}$ &  &   0.7 &   9.4 &  18.6 &  &   0.7 &   8.6 &  17.4 &  &   0.0 &   2.0 &   8.3 \\ 
$15p^{C0.5}$ &  &   0.8 &  11.4 &  20.1 &  &   1.1 &  11.9 &  20.3 &  &   0.0 &   3.4 &  13.1 \\ 
$15p^{C0.9}$ &  &   0.9 &  12.7 &  20.7 &  &   1.3 &  14.3 &  22.5 &  &   0.0 &   3.7 &  16.4 \\ 
$15p^{M0.2}$ &  &   0.7 &   9.5 &  18.7 &  &   0.7 &   8.8 &  17.6 &  &   0.0 &   2.0 &  11.9 \\ 
$15p^{M0.5}$ &  &   0.8 &  11.4 &  19.9 &  &   1.0 &  12.1 &  20.5 &  &   0.0 &   2.8 &  12.5 \\ 
$15p^{M0.9}$ &  &   0.9 &  12.1 &  20.4 &  &   1.1 &  13.1 &  22.0 &  &   0.0 &   3.4 &  15.5 \\ 
$16i$ &  &   0.9 &   6.4 &  56.8 &  &   0.9 &   4.3 &  15.7 &  &   1.7 &   2.0 &  45.1 \\ 
$16i^{C0.2}$ &  &   0.7 &   3.3 &  39.2 &  &   0.6 &   2.2 &   6.8 &  &   0.0 &   1.9 &  37.7 \\ 
$16i^{C0.5}$ &  &   0.8 &   4.4 &  50.6 &  &   0.6 &   2.3 &  10.6 &  &   0.0 &   1.9 &  40.3 \\ 
$16i^{C0.9}$ &  &   0.8 &   5.1 &  53.9 &  &   0.6 &   3.0 &  12.8 &  &   0.0 &   1.9 &  42.7 \\ 
$16i^{M0.2}$ &  &   0.7 &   3.5 &  41.2 &  &   0.6 &   2.2 &   7.9 &  &   1.5 &   1.9 &  37.9 \\ 
$16i^{M0.5}$ &  &   0.8 &   4.3 &  42.0 &  &   0.6 &   2.3 &   9.9 &  &   1.5 &   1.9 &  41.4 \\ 
$16i^{M0.9}$ &  &   0.8 &   4.6 &  53.2 &  &   0.6 &   2.5 &  12.3 &  &   1.5 &   1.9 &  40.1 \\ 
$16p$ &  &   1.4 &   8.4 &  20.7 &  &   1.4 &   7.2 &  13.1 &  &   0.0 &   2.1 &   4.4 \\ 
$16p^{C0.2}$ &  &   1.0 &   4.4 &  13.8 &  &   0.7 &   2.8 &   6.5 &  &   0.0 &   1.7 &   2.3 \\ 
$16p^{C0.5}$ &  &   1.1 &   5.9 &  15.9 &  &   0.8 &   4.2 &   8.4 &  &   0.0 &   1.7 &   2.3 \\ 
$16p^{C0.9}$ &  &   1.2 &   6.5 &  18.0 &  &   1.0 &   5.1 &  10.2 &  &   0.0 &   1.8 &   3.2 \\ 
$16p^{M0.2}$ &  &   1.0 &   4.5 &  13.9 &  &   0.7 &   3.0 &   6.7 &  &   0.0 &   1.7 &   2.3 \\ 
$16p^{M0.5}$ &  &   1.1 &   5.9 &  16.7 &  &   0.8 &   4.2 &   8.6 &  &   0.0 &   1.7 &   2.3 \\ 
$16p^{M0.9}$ &  &   1.2 &   6.2 &  16.9 &  &   0.9 &   4.5 &   9.1 &  &   0.0 &   1.8 &   3.2 \\ 
$17i$ &  &   1.8 &  18.3 &  81.3 &  &   1.8 &  15.5 &  53.1 &  &   1.7 &   5.0 & 110.4 \\ 
$17i^{C0.2}$ &  &   1.3 &   9.0 &  34.1 &  &   1.0 &   4.2 &  14.8 &  &   0.0 &   2.2 &  99.7 \\ 
$17i^{C0.5}$ &  &   1.5 &  12.6 &  41.6 &  &   1.2 &   6.9 &  21.8 &  &   0.0 &   3.0 & 106.5 \\ 
$17i^{C0.9}$ &  &   1.6 &  15.0 &  45.3 &  &   1.4 &   9.7 &  26.6 &  &   0.0 &   4.4 & 108.8 \\ 
$17i^{M0.2}$ &  &   1.3 &   9.0 &  32.6 &  &   1.1 &   4.4 &  14.6 &  &   1.5 &   2.1 & 104.2 \\ 
$17i^{M0.5}$ &  &   1.5 &  13.2 &  43.7 &  &   1.2 &   6.8 &  22.5 &  &   1.5 &   3.6 & 102.0 \\ 
$17i^{M0.9}$ &  &   1.5 &  14.0 &  42.6 &  &   1.3 &   8.3 &  24.8 &  &   1.5 &   4.2 & 108.8 \\ 
$17p$ &  &   1.1 &  11.4 &  22.0 &  &   2.1 &  15.9 &  24.7 &  &   0.0 &   5.7 &  18.6 \\ 
$17p^{C0.2}$ &  &   0.7 &   6.9 &  16.1 &  &   1.0 &   6.0 &  13.7 &  &   0.0 &   2.0 &   6.0 \\ 
$17p^{C0.5}$ &  &   0.9 &   8.8 &  18.2 &  &   1.4 &   8.6 &  16.3 &  &   0.0 &   3.0 &  10.6 \\ 
$17p^{C0.9}$ &  &   1.0 &  10.1 &  19.4 &  &   1.7 &  11.0 &  18.1 &  &   0.0 &   3.6 &  15.1 \\ 
$17p^{M0.2}$ &  &   0.7 &   6.9 &  16.1 &  &   1.0 &   6.2 &  13.7 &  &   0.0 &   2.0 &   8.2 \\ 
$17p^{M0.5}$ &  &   0.9 &   8.9 &  18.4 &  &   1.4 &   8.8 &  17.2 &  &   0.0 &   3.3 &  12.7 \\ 
$17p^{M0.9}$ &  &   1.0 &   9.4 &  18.8 &  &   1.5 &   9.6 &  17.5 &  &   0.0 &   3.3 &  15.0 \\ 
$18i$ &  &   2.2 &  39.7 & 124.6 &  &   3.9 &  39.6 & 117.2 &  &   1.7 &   8.8 &  62.2 \\ 
$18i^{C0.2}$ &  &   1.3 &  13.9 &  92.7 &  &   1.4 &   9.0 &  90.8 &  &   0.0 &   3.3 &  49.0 \\ 
$18i^{C0.5}$ &  &   1.7 &  21.4 &  95.6 &  &   1.9 &  15.5 &  91.1 &  &   0.0 &   5.3 &  53.4 \\ 
$18i^{C0.9}$ &  &   1.9 &  26.2 &  96.8 &  &   2.2 &  21.8 &  92.0 &  &   0.0 &   7.5 &  55.2 \\ 
$18i^{M0.2}$ &  &   1.3 &  14.8 &  92.5 &  &   1.3 &   9.4 &  90.5 &  &   1.5 &   3.1 &  46.1 \\ 
$18i^{M0.5}$ &  &   1.6 &  21.2 &  96.6 &  &   1.8 &  17.1 &  91.1 &  &   1.5 &   5.5 &  55.2 \\ 
$18i^{M0.9}$ &  &   1.8 &  23.0 &  95.8 &  &   2.0 &  18.6 &  91.3 &  &   1.5 &   7.0 &  53.7 \\ 
$18p$ &  &   0.9 &  18.0 &  27.2 &  &   0.1 &  20.7 &  32.7 &  &   0.0 &  13.8 &  30.1 \\ 
$18p^{C0.2}$ &  &   0.9 &  12.3 &  22.4 &  &   1.3 &   9.9 &  22.9 &  &   0.0 &   4.0 &   8.7 \\ 
$18p^{C0.5}$ &  &   0.9 &  15.3 &  24.6 &  &   1.3 &  13.4 &  26.8 &  &   0.0 &   6.8 &  13.4 \\ 
$18p^{C0.9}$ &  &   0.9 &  16.9 &  26.0 &  &   1.3 &  15.6 &  29.7 &  &   0.0 &   8.9 &  19.8 \\ 
$18p^{M0.2}$ &  &   0.9 &  12.1 &  22.5 &  &   1.1 &  10.2 &  24.0 &  &   0.0 &   4.4 &  10.1 \\ 
$18p^{M0.5}$ &  &   0.9 &  14.9 &  24.8 &  &   1.1 &  13.4 &  27.9 &  &   0.0 &   6.5 &  17.5 \\ 
$18p^{M0.9}$ &  &   0.9 &  15.8 &  25.1 &  &   1.1 &  14.9 &  29.4 &  &   0.0 &   8.6 &  15.2 \\ 
$19i$ &  &  25.1 &  73.0 & 140.3 &  &  26.7 &  78.0 & 123.5 &  &  26.5 & 115.3 & 119.1 \\ 
$19i^{C0.2}$ &  &  10.2 &  43.2 & 104.9 &  &   6.3 &  34.0 &  91.3 &  &   5.9 &  25.0 & 113.2 \\ 
$19i^{C0.5}$ &  &  15.2 &  53.2 & 116.1 &  &  11.6 &  50.6 &  92.6 &  &  11.3 &  36.6 & 117.1 \\ 
$19i^{C0.9}$ &  &  18.4 &  58.4 & 136.6 &  &  16.0 &  58.9 &  95.8 &  &  16.9 &  41.2 & 118.0 \\ 
$19i^{M0.2}$ &  &  11.0 &  43.5 & 105.0 &  &   6.4 &  35.2 &  91.3 &  &   8.2 &  31.9 & 112.6 \\ 
$19i^{M0.5}$ &  &  15.3 &  52.2 & 118.4 &  &  11.6 &  50.8 &  93.4 &  &  12.8 &  41.2 & 118.0 \\ 
$19i^{M0.9}$ &  &  16.7 &  55.7 & 134.0 &  &  13.3 &  55.3 &  94.3 &  &  13.1 &  41.2 & 115.4 \\ 
 
\enddata
\end{deluxetable}

\end{document}